\documentclass{rspublic}
\newcommand{\abs}[1]{\lvert#1\rvert}
\newcommand{\norm}[1]{\lVert#1\rVert}
\newcommand{\dom}[1]{\ensuremath{\mathcal{D}(\opr{#1})}}
\newcommand{\opr}[1]{\ensuremath{\mathbf{\mathsf{#1}}}}
\providecommand{\mcal}{\ensuremath{\mathcal}}
\begin{document}

\title[Pauli's theorem and quantum canonical pairs]{Pauli's Theorem and Quantum Canonical Pairs:\\The Consistency Of a Bounded, \\Self-Adjoint Time Operator Canonically\\ Conjugate to a  Hamiltonian with Non-empty Point Spectrum}

\author[E.A. Galapon]{Eric A. Galapon}

\affiliation{Theoretical Physics Group\\National Institute of Physics, University of the Philippines\\Diliman, Quezon City, Philippines 1101}
\maketitle
\begin{abstract}{Pauli's theorem; quantum canonical pairs; time operators} 
In single Hilbert space, Pauli's well-known theorem implies that the existence of a self-adjoint time operator canonically conjugate to a given Hamiltonian signifies that the time operator and the Hamiltonian possess completely continuous spectra spanning the entire real line.  Thus the conclusion that there exists no self-adjoint time operator conjugate to a semibounded or discrete Hamiltonian despite some well-known illustrative counterexamples.  In this paper we evaluate Pauli's theorem against the single Hilbert space formulation of quantum mechanics, and consequently show the consistency of assuming a bounded, self-adjoint time operator canonically conjugate to a Hamiltonian with an unbounded, or semibounded, or finite point spectrum.  We point out Pauli's implicit assumptions and show that they are not consistent in a single Hilbert space. We demonstrate our analysis by giving two explicit examples. Moreover, we clarify issues sorrounding the different solutions to the canonical commutation relations, and, consequently, expand the class of acceptable canonical pairs beyond the solutions required by Pauli's theorem.
\end{abstract}
\section{Introduction}

Despite of the immense success of the standard single Hilbert space formulation of quantum mechanics, many remain unsatisfied with the formulation because of its failure to accommodate some non-trivial problems such as the problem of incorporating a dynamical theory of time in quantum mechanics. In standard quantum mechanics, time is a mere parameter, an external variable independent from the dynamics of any given system (Omnes 1994).  But time undoubtedly acquires dynamical significance in questions involving the occurrence of an event (Busch 1990$a$; Srinivas \& Vijayalakshmi 1981), e.g. when a nucleon decays (Eisenberg \& Horwitz 1997; Peres 1980), or when a particle arrives at a given spatial point (Muga, Sala \& Palo 1998; Grot {\it et. al.} 1996), or when a particle emmerges from a potential barrier (Landauer 1994).  Furthermore, the time-energy uncertainty principle requires more than a parametric treatment of time (Aharonov \& Bohm 1969; Busch 1990$a$,$b$). Standard quantum mechanics requires a self-adjoint time operator conjugate to the Hamiltonian for each of the above mentioned examples.  However, with the Hamiltonian generally possessing a semi-bounded, pure point spectrum, finding self-adjoint time operators has been deemed impossible to achieve (Toller 1997, 1999; Giannitrapani 1997;  Delgado \& Muga 1997; Park 1984; Holevo 1978; Cohen-Tannoudji 1977; Olhovsky \& Recami 1974; Gottfried 1966; Pauli 1926).  

The reason for this pessimism is a theorem due to Pauli (1926, 1933, 1958) which has been believed to assert without exception that there exists no self-adjoint time operator canonically conjugate to a semi-bounded Hamiltonian in single Hilbert space: The existence of a self-adjoint time operator implies that the time operator and the Hamiltonian have completely continuous spectra spanning the entire real line, or, in modern parlance, the time operator and the Hamiltonian form a system of imprimitivities based on the real line $\Re$. While some examples of self-adjoint canonical pairs are known that do not possess spectra spanning $\Re$---the momentum and the position operators of a particle trapped in a box (Reed \& Simon 1975; Segal 1967; Nelson 1959), the angular momentum and the angle operators (Kraus 1965), the harmonic oscillator number and phase operators (Galindo 1984, Garrison \&Wong 1970)---Pauli's Theorem remains unquestioned and continues to be a major motivation in shaping much of the present works in incorporating a dynamical theory of time in quantum mechanics (see for example Kuusk \& Koiv 2001 for a recent reference to Pauli's theorem). This has led to diverse treatments of time within (Leon {\it et. al.} 2000; Giannitrapani 1997; Busch {\it et. al} 1994, 1995$a$,$b$; Holevo 1978; Helstrom 1970, 1976) and beyond the usual formulation and interpretation of quantum mechanics ( Eisenberg \& Horwitz 1997; Halliwell \& Zafiris 1997; Blanchard \& Jadczk 1996; Holland 1993;  Busch 1990$a$,$b$; Rosenbaum 1969).

In fairness to Pauli, he forwarded his well-known theorem at a time when von Neumann's geometric reformulation of quantum mechanics--the now standard single Hilbert space formulation--was not yet in place. However, after von Neumann's work (von Neumann 1932, 1955), it has been the assumption of the majority that Pauli's argument, together with its attending conclusion, remains valid in the single Hilbert space formulation, thus the prevalent belief that no self-adjoint time operator can be constructed in quantum mechanics (see, for example, Jammer 1974). It is then the objective of this paper to evaluate the validity of the objections raised by Pauli within the standard single Hilbert space formulation of quantum mechanics.  Specifically, we will rigorously show  the consistency of assuming the existence of a self-adjoint, bounded time operator canonically conjugate to a  Hamiltonian with a non-empty unbounded, semibounded, or finitely countable point spectrum. Our method of proof retraces the steps followed by Pauli leading to his conclusion. In the process, we will uncover the implicit assumptions made by Pauli and show that these assumptions are not mutually consistent under the assumption that the time operator is bounded. While we specifically deal with bounded time operators, our results and conclusion are sufficient to point out that Pauli's Theorem, together with its attending conclusion, does not hold in single Hilbert space quantum mechanics. In addition to our evaluation of Pauli's argument, we attempt to clarify the issue sorrounding the different solutions---solutions with distinct properties such as spectral properties---to the canonical commutation relation. We will argue that imprimitivity based on the real line (condition arising from Pauli's theorem as applied in single Hilbert spaces) is not an esteemed property that has to be imposed on all physically acceptable canonical pairs.

The paper is organized as follows.  In Section-\ref{revisit} we revisit Pauli's well known Theorem.  In Section-\ref{consistency} we prove four theorems, which we shall refer to as the consistency theorems, to show the consistency of assuming a self-adjoint, bounded time operator conjugate to Hamiltonian with a non-empty point spectrum.  In Section-\ref{synthesis} we synthesize the consistency theorems and there point out the inconsistent assumptions made by Pauli.  In Section-\ref{explicit}  we give an example explicitly demonstrating the consistency theorems; similarly in Section-\ref{free} we give an example of a bounded, self-adjoint operator canonically conjugate to the semibounded and discrete Hamiltonian of a confined particle. In Section-\ref{discussion} we clarify the confusion surrounding Pauli's Theorem. And in Section-\ref{canonical} we discuss the physical relevance of the different solutions to the canonical commutation relation. In the following, whenever we refer to Pauli's theorem we shall mean the theorem against the backdrop of a single Hilbert space.  All through out we set $\hbar=1$.

\section{Revisiting Pauli's Theorem}\label{revisit}

In single Hilbert space language, Pauli's argument  goes as follows:  Let $H$ be a self-adjoint Hamiltonian. Assume that there exists a self-adjoint time operator,$T$, canonically conjugate to $H$, i.e. 
\begin{equation} \label{ccrel} 
\left[T,H\right]=i     
\end{equation}
Since $T$ is self-adjoint, then for all $\beta\in\Re$, $U_\beta=\exp(-i\beta T)$ is unitary. A formal expansion of the exponential yields the commutator
\begin{equation} \label{paul} 
\left[U_\beta, H \right] = 
\sum_{k=0}^\infty \frac{\left(-i \beta \right)^{k}}{k!}\left[T^k , H \right]=-\beta U_\beta \end{equation}        
in which use is made of the commutation relation (\ref{ccrel}).  Let $\varphi_E$ be an eigenvector of $H$ with eigenvalue $E$.  It is a direct consequence of the commutation relation (\ref{paul}) that
\begin{equation} \label{conc} 
H U_\beta \varphi_E=\left(E+\beta\right) U_\beta \varphi_E. 
\end{equation}
That is $U_\beta \varphi_E$ is an eigenvector of $H$ with the eigenvalue $\left(E+\beta \right)$.  This implies that $H$ has a continuous spectrum spanning the entire real line because $\beta$ is an arbitrary real number. Equation (\ref{conc}) then asserts that no self-adjoint time operator exists that is canonically conjugate to the generally semibounded and discrete Hamiltonian of quantum mechanics; otherwise, the supposedly unitary   will map the discrete or bounded spectrum of H into the entire real line, which is not possible. Thus Pauli wrote, ``We conclude that the introduction of an operator $T$ must fundamentally be abandoned and that the time $t$ in quantum mechanics has to be regarded as an ordinary number.''   

However, the above proof is formal, i.e. without regard to the domains of the operators involved and to the validity of the operations leading to the conclusion.  Equation (\ref{conc}) follows from two implicit assumptions that for arbitrary real number $\beta$ (i) the formal expansion in equation (\ref{paul}) is valid, and (ii) the operation $\left[U_\beta , H \right] \varphi_E=\beta U_\beta \varphi_E$ makes sense.  (In the original formulation of the theorem, Hemitian was used for self-adjoint; both mean the same thing within the context of the original theorem.) It is only when these assumptions are consistent that the contradiction raised by Pauli is legitimate. Otherwise the conclusion does not hold. In the following section we will show in a single Hilbert space that while assuming a bounded self-adjoint time operator to ensure the validity of the former assumption the later does not hold.

\section{The Consistency Theorems}\label{consistency}

The range and implications of Pauli's theorem are not only limited to the time aspect of quantum mechanics, but also to other aspects such as the angular momentum and the harmonic oscillator phase problems.  So in the following we extend our discussion to a general pair, $(Q,P)$, and treat the time-operator problem as a special case.  We shall take the standard single Hilbert space formulation of quantum mechanics as our reference. In the standard formulation, to every system is attached a separable Hilbert space over the complex field; and in the absence of superselection rules, every ray corresponds to a physical state; and every linear, densely defined, self-adjoint operator acting on the Hilbet space corresponds to a physical obsevable. The definition of a canonical pair given below, Definition-\ref{defccp}, is the most general definition that can accommodate whatever consistent unstated assumptions appear in the definition of a canonical pair in Pauli's theorem.

We shall proceed in parallel to Pauli's Theorem and proof. We assume the existence of a pair of self-adjoint canonically conjugate operators in a not necessarily dense subspace of the Hilbert space involved; we further assume that one of the pair, specifically $Q$, is bounded (as implicitly assumed by Delgado and Muga (1997) for the time operator). We then proceed in deducing the consequences of these assumptions.    Our goal is to arrive at equation (\ref{paul}) and show that no contradiction arises.  Basically we will give answer to the question: With $Q$ (time operator) bounded (thus assuring that the expansion in equation (\ref{paul}) is valid for arbitrary real number $\beta$) under what conditions equation (\ref{paul}) holds in the entire domain of $P$ (the Hamiltonian)?  

In the following we prove the consistency of the above assumptions without succumbing to Pauli's conclusion. The proof consists of four theorems, which we shall refer to as the consistency theorems. We state and prove the theorems without comment on their relevance to Pauli's argument.  We will give a synthesis of the theorems in the following section and there point out the inconsistent assumptions leading to equation (\ref{conc}). In the following all operators are linear in keeping with standard quantum mechanics.

\begin{definition} A not necessarily densely defined operator $C$ with domain $\mathcal{D}(C)$ in an infinite dimensional, separable Hilbert space, $\mathcal{H}$, is a commutator if there exists a pair of densely defined operators $A$ and $B$ in the same Hilbert space such that $(AB-BA)\varphi=C\varphi$ for all $\varphi \in \mathcal{D}(C) \subseteq \mathcal{D}(AB) \cap \mathcal{D}(BA)$.  We shall denote a commutator by the triple $C(A,B; \mathcal{D}(C))$. \end{definition}

\begin{remark} We emphasize that a commutator $C(A,B; \mathcal{D}(C))$ always means that the relation $(AB-BA)\varphi=C\varphi$ holds strictly in the domain $\mathcal{D}(C)$.
\end{remark}

\begin{lemma} \label{inclu} Let $C$ be the commutator $C(A,B; \mathcal{D}(C))$.  Suppose $A$ ($B$) is self-adjoint with a non-empty point spectrum. A necessary condition for the eigenvectors of $A$ ($B$) to belong to $\mathcal{D}(C)$ is that $C$ maps each of the eigenvectors of $A$ ($B$) to its orthogonal complement. \end{lemma}

\begin{proof} Let $A \varphi = a \varphi$ and $\varphi \in \mathcal{D}(C)$. Because the eigenvalue $a$ is real and $\abs{\left<\varphi \big{|} B \varphi \right>}< \infty$, we have the equality $\left< \varphi \big{|} C \varphi \right> = \left<\varphi \big{|} (AB-BA) \varphi \right>=0$.  That is $\left< \varphi \big{|} C \varphi \right> =0$, or $\varphi$ is orthogonal to $C\varphi$. \end{proof}

\begin{definition} We refer to the commutator $C(A,B; \mathcal{D}(C))$ as an invariance commutator if $\mathcal{D}(C)$ is invariant under $C$ ; that is, $C \mathcal{D}(C)\subseteq\mathcal{D}(C)$, or for every $\phi\in\mathcal{D}(C)$, $C\phi\in \mathcal{D}(C)$. \end{definition}

\begin{lemma} \label{cmlem} Let $C$ be the invariance commutator $C(A,B; \mathcal{D}(C))$.  Let $C^{-1}$ exists. If $C$ and $C^{-1}$ are both bounded, and if $\mathcal{D}(C)$ is invariant under $A$ ($B$), and if $C$ commutes with $A$ ($B$) in $\mathcal{D}(C)$, i.e. $AC \varphi =CA \varphi$ ($BC \varphi =CB \varphi$)  for all $\varphi \in \mathcal{D}(C)$, then $A$ and $B$ can not be simultenously bounded and defined every where. \end{lemma}

\begin{proof} Suppose that $A$ and $B$ are bounded and defined everywhere so that any algebraic expression involving them is defined everywhere. Moreover assume that $\mathcal{D}(C)$ is invariant under $A$ and that $A^n \neq 0$ for arbitrary positive integer $n$. By assumption all the operators involved are linear; this implies that $\mathcal{D}(C)$ is it self linear. By the linearity of $\mathcal{D}(C)$, we have $(A^2B-BA^2)\varphi=A(AB-BA)\varphi+(AB-BA)A\varphi$. The assumed invariance of $\mathcal{D}_c$ under $A$ and the commutativity of $A$ and $C$ in $\mathcal{D}(C)$ yield $(A^2B-BA^2)\varphi=AC\varphi+CA\varphi=2AC\varphi$. Then by induction we get the expression
\begin{equation}\label{induc}
(A^n B -B A^n) \varphi = n A^{n-1} C \varphi 
\end{equation}
\noindent for all $\varphi \in \mathcal{D}(C)$ and for every positive integer $n$.  

Taking the norm of both sides of equation (\ref{induc}) gives the inequality 
\begin{equation} \label{inequa}
	n \norm{A^{n-1} C \varphi} \le \norm{A^n B \varphi}+\norm{B A^n \varphi} 
\end{equation}

\noindent  Now let us restrict equation (\ref{inequa}) in $\mathcal{R}(C)\subseteq\mathcal{D}(C)$.
Dividing both sides of equation (\ref{inequa}) by $\norm{\varphi} \neq 0$ and taking the supremum in $\mathcal{R}(C)$,
\begin{eqnarray} 
	n \sup_{\substack{\varphi \in \mathcal{R}(C) \\ 
	\varphi\neq0}} \frac{\norm{A^{n-1}C\varphi}}{\norm{\varphi}} &\le&  		\sup_{\substack{\varphi \in \mathcal{R}(C) \\ 
	\varphi \neq 0}} \frac{\norm{A^nB\varphi}}{\norm{\varphi}} + \sup_{\substack{\varphi \in 		\mathcal{R}(C)\\ 
	\varphi \neq 0}} \frac{\norm{BA^n \varphi}}{\norm{\varphi}}  \nonumber \\      	n\norm{A^{n-1}C}_C &\le& 2 \norm{A^n}_C \norm{B}_C \nonumber \\
	&\le& 2\norm{A^{n-1}}_C \norm{A}_C \norm{B}_C \nonumber \\
	&=& 2\norm{A^{n-1} C C^{-1}}_C \norm{A}_C \norm{B}_C \label{inin} \\
	&\le& 2 \norm{A^{n-1} C}_C \norm{C^{-1}}_C \norm{A}_C \norm{B}_C \label{ineq}
\end{eqnarray}
where $\norm{\cdot}_C$ is the restricted norm in $\mathcal{R}(C)$.  The restriction above in $\mathcal{R}(C)$ is necessary to enable us to have the equality in equation (\ref{inin}).  Cancelling the common factors in (\ref{ineq}) finally leads to the inequality
\begin{equation} \label{finineq}
	n \le 2 \norm{A}_C \norm{C^{-1}}_C \norm{B}_C 
\end{equation}
which is impossible for arbitrary integer $n$ because $A$, $C^{-1}$ and $B$ are bounded, i.e. have finite norms.  

If $A$ is nilpotent of index $n \ge2$, then equation (\ref{induc}) implies that $n A^{n-1} C \varphi=0$ for all $\varphi\in\mathcal{D}(C)$, which is also impossible for $C \ne 0$.  Thus $A$ and $B$ can not be simultenously bounded and defined everywhere.  The same conclusion can be arrived by interchanging the roles of $A$ and $B$. \end{proof}                                                                 

\begin{remark}When $\mathcal{D}(C)$ coincides with the entire Hilbert space, the condition on the invariance of $\mathcal{D}(C)$ under either $A$ or $B$ becomes unnecessary. The  condition that $C$ is an invariance commutator can be replaced by the weaker condition that there is a non-trivial intersection of $\mathcal{D}(C)$ and  $\mathcal{R}(C)$. Under this condition we arrive at an inequality similar to (\ref{finineq}) by restricting equation (\ref{induc}) in $\mathcal{D}(C)\cap\mathcal{R}(C)$. \end{remark}

\begin{definition}\label{defccp} We call a pair of self-adjoint operators, $Q$ and $P$, in an infinite dimensional, separable Hilbert space, $\mathcal{H}$, a canonical pair if 
\begin{equation}\label{ccr} (QP -P Q)\phi=i\phi\; \mbox{for all}\;\phi \in \mathcal{D}_c
\end{equation}
where $\mathcal{D}_c$ is a dense or closed proper subspace of $\mathcal{H}$. \end{definition}

\begin{theorem} \label{ccpair}Let $Q$  and $P$  be a canonical pair in the domain $\mathcal{D}_c$ .  (i) Then the eigenvectors of $Q$ and $P$, if they exist, do not belong to $\mathcal{D}_c$. (ii) If $\mathcal{D}_c$ is invariant under either $Q$ or $P$, then $Q$ and $P$  can not be both bounded. \end{theorem}

\begin{proof} (i) The canonical pair $Q$ and $P$, and the domain $\mathcal{D}_c$ constitute the commutator $C=iI_c$ where $I_c$ is the identity in $\mathcal{D}_c$.  Assuming that $Q$ and $P$ have eigenvectors in ${\cal D}_c$, their eigenspaces are obviously invariant under $C=iI_c$.  Thus by Lemma~\ref{inclu} the eigenvectors do not belong to $\mathcal{D}_c$. (ii)  If $\mathcal{D}_c$ is invariant under either $Q$ or $P$, then it follows immediately from Lemma~\ref{cmlem} that $Q$ and $P$ can not be both bounded because $C=iI_c$ is an invariance commutator, and it has the well defined inverse $C^{-1}=-iI_c$, and $C$ obviously commutes with any operator which leaves $\mathcal{D}_c$ invariant. \end{proof}

\begin{remark}
When $\mcal{D}_c$ is dense, the condition on the invariance of $\mcal{D}_c$ under either $Q$ or $P$ may not be necessary; this is the well-known Wielandt's Theorem. However, when $\mcal{D}_c$ is closed, the invariance condition is necessary. Without this condition, it is possible to construct a pair of bounded and self-adjoint operators satisfying the canonical commutation relation in a closed subspace of the Hilbert space. Consider the following pair of everywhere defined, self-adjoint operators in the Hilbert space $\mcal{H}=L^2[0,1]$,
\begin{equation}\label{p1}
\left(Q\phi\right)\!(q)=q\phi(q)
\end{equation}
\begin{equation}\label{p2}
\left(P\phi\right)\!(q)=\int_0^1 \frac{3\sqrt{5}}{2}\left(q+q'\right)\phi(q')\,dq'.
\end{equation}
It is straightforward to show that the pair satisfies equation (\ref{ccr}) in the following closed, one-dimensional subspace,
\begin{equation}
\mcal{D}_c=\left\{c\left(i q^2-\frac{2\sqrt{5}}{15}-\frac{i}{3}\right),\,c\in\mcal{C}\right\}.
\end{equation}
Obviously $\mcal{D}_c$ is not invariant under either $Q$ or $P$. 
\end{remark}

\begin{theorem} \label{comain} Let $Q$ and $P$ be a canonical pair in the domain $\mathcal{D}_c$. Let $P$ be unbounded. And let $\mathcal{D}_c$ be invariant under $Q$.  (i) If $Q$ is bounded and defined everywhere, and the restriction of $P$ in $\mathcal{D}_c$ is closed, then in the strong operator topology
\begin{equation}
            \label{paul-Dc}      (U_\beta P-PU_\beta)\varphi = \beta U_\beta \varphi 
\end{equation}
for all $\varphi \in \mathcal{D}_c$ and $\beta \in \Re$ where $U_\beta = e^{-i \beta Q}$.  (ii) Moreover, if $\mathcal{D}_c$ is dense and if there exists a $\beta' \in \Re$ such $\mathcal{D}(P)\setminus\mathcal{D}_c$ is invariant under $U_{\beta'}$, i.e. for every $\phi\in\mathcal{D}(P)\setminus\mathcal{D}_c$, $U_{\beta'}\phi \in \mathcal{D}(P)\setminus\mathcal{D}_c$, then equation (\ref{paul-Dc}) holds in $\mathcal{D}(P)\setminus\mathcal{D}_c$. That is 
\begin{equation} \label{paul-Dp}  
(U_{\beta'}P-P U_{\beta'}) \phi=\beta' U_{\beta'} \phi 
\end{equation}
for all $\phi \in \mathcal{D}(P)\setminus\mathcal{D}_c$.  \end{theorem}

\begin{proof} (i) Consider the following sequence of bounded, everywhere defined operators, 
$U_\beta^{(n)}=\sum_{k=0}^n (k!)^{-1}(-i\beta)^k Q^k$, for $n=0,1,2, \ldots$, for all $\beta \in \Re$. Since $\mathcal{D}_c$ is invariant under $Q$, we have $P U_\beta^{(n)} \varphi = \sum_{k=0}^n (k!)^{-1}(-i \beta)^k P Q^k \varphi$ for all $\varphi \in D_c$.   The invariance of $\mathcal{D}_c$ under $Q$ and the boundedness of $Q$ ensure that $\mathcal{D}(PQ^k)\supseteq\mathcal{D}_c$ and $\mathcal{D}(Q^kP)\supseteq\mathcal{D}_c$; therefore $(Q^k P-PQ^k)$ is defined in the entire $\mathcal{D}_c$.  The linearity of $\mathcal{D}_c$ leads to $(Q^k P-PQ^k) \varphi=ikQ^{k-1} \varphi$ for all $\varphi \in D_c$ and for every positive integer $k\prec \infty$.  Using this result we get
\begin{equation} 
	\label{inter}    (U_\beta^{(n)} P-P U_\beta^{(n)})\varphi=\beta U_\beta^{(n-1)} \varphi \end{equation}
To establish commutation relation (\ref{paul-Dc}) we need to get the limiting form of eqn.(\ref{inter}). Note that $U_\beta^{(n)} P \varphi$  and $U_\beta^{(n)}\varphi$ are strong Cauchy sequences converging to $U_\beta P \varphi$ and $U_\beta \varphi$, respectively.  Also the sequence $PU_\beta^{(n)} \varphi$ is strongly Cauchy.  This follows from equation (\ref{inter}), $PU_\beta^{(n)} \varphi=\beta U_\beta^{(n-1)} \varphi - U_\beta^{(n)} P \varphi$.  Because the two terms in the right hand side of the equation are themselves strong Cauchy sequences, then by an $\epsilon/2$ argument $P U_\beta^{(n)} \varphi$ is itself a strong Cauchy sequence converging to some $\xi$. Since the restriction of $P$ in $\mathcal{D}_c$ is closed, and since $U_\beta^{(n)}\varphi \in\mathcal{D}_c$ for all $n$ and is strongly convergent, then $PU_\beta^{(n)}\varphi \rightarrow PU_\beta \varphi$ and $U_\beta \varphi \in\mathcal{D}_c$. Thus in the limit
\begin{displaymath} 
	(U_\beta P-P U_\beta) \varphi=\beta U_\beta \varphi 
\end{displaymath}

\noindent for all $\varphi \in \Re$, $\beta \in \Re$, which is what we have sought to prove.  

(ii) Let $\phi \in \mathcal{D}(P)\backslash\mathcal{D}_c$ and $\varphi\in\mathcal{D}_c$.  For the given $\beta'$, we have, by the first half of the theorem (on using the conjugate relation of equation (\ref{paul-Dc})), 
\begin{displaymath}
\left<\phi \big{|}( P U_{\beta'}^{\ast}-U_{\beta'}^{\ast} P)\varphi \right>=\beta' \left<\phi \big{|} U_{\beta'}^{\ast} \varphi \right>  
\end{displaymath}

\noindent Because $\mathcal{D}(P)\setminus\mathcal{D}_c$ is invariant under $U_{\beta'}$ by assumption, we have 
\begin{displaymath} 
\left<(U_{\beta'} P-P U_{\beta'}) \phi \big{|} \varphi \right> = \beta' \left<U_{\beta'} \phi \big{|} \varphi \right> \rightarrow  \left<(U_{\beta'}P-PU_{\beta'}) \phi - \beta' U_{\beta'} \phi \big{|} \varphi \right>=0. 
\end{displaymath}

\noindent Since $\phi$ and $\varphi$ are arbitrary and $\mathcal{D}_c$ is dense, i.e. orthogonal to $0$, we finally have
\begin{displaymath} 
(U_{\beta'} P-P U_{\beta'}) \phi=\beta' U_{\beta'} \phi 
\end{displaymath}

\noindent for all $\phi \in \mathcal{D}(P)\setminus\mathcal{D}_c$.  
\end{proof}

\begin{remark}  It is possible that only a subspace of $\mathcal{D}_c$, say $\mathcal{D}$, is invariant under the action of $Q$.  For this case, if all conditions in theorem-\ref{comain} are satisfied in $\mathcal{D}$, then  theorem-\ref{comain} holds in $\mathcal{D}$ and in $\mathcal{D}(P)\setminus\mathcal{D}$.
\end{remark} 

\begin{definition} Let $B_I$ be the set of all points $\beta \in \Re$ such that $\mathcal{D}(P)\setminus\mathcal{D}_c$ is invariant under the unitary operator $U_{\beta}=e^{-i\beta Q}$.  We refer to $B_I$ as the invariance parameter set.
\end{definition}

\begin{theorem} \label{main} Let all conditions be satisfied in theorem-\ref{comain}.  Moreover, let the point spectrum of $P$ and the invariance parameter set $B_I$ be non-empty. (i) If $\beta_0\in B_I$, then $n\beta_0 \in B_I$ for all integer $n=\pm1, \pm2, \ldots$. (ii) If $\varphi_0$ is a normalized eigenvector of $P$ with eigenvalue $p_0$, then the vectors $\varphi_n=U_{\beta_0}^n \varphi_0$, $n=\pm1, \pm2, \ldots$, in which $U_{\beta_0}^{-\abs{n}} \varphi_0=U_{-\beta_0}^{\abs{n}} \varphi_0$, are normalized eigenvectors of $P$ with eigenvalues $p_n=(p_0+n\beta_0)$, respectively.  (iii) Furthermore if $p_0$ has a multiplicity $\lambda$ then the rest of the eigenvalues have the same multiplicity $\lambda$.  (iv) $P$ has infinitely countable eigenvalues extending from negative infinity to positive infinity, and (v) $B_I$ consists of the differences of these eigenvalues. 
\end{theorem}

\begin{proof} (i) Let $\phi$ be an arbitrary element of  $\mathcal{D}(P)\setminus\mathcal{D}_c$. Then $U_{\beta_0}^2\phi = U_{\beta_0}\cdot U_{\beta_0}\phi = U_{\beta_0}\psi_1=\psi_2 \in \mathcal{D}(P)\setminus\mathcal{D}_c$ because $\psi_1=U_{\beta_0}\phi \in \mathcal{D}(P)\setminus\mathcal{D}_c$. Then, by induction, for every $\phi \in \mathcal{D}(P)\setminus\mathcal{D}_c$  and for every positive integer $n$, $U_{\beta_0}^n \phi=U_{n\beta_0}\phi \in \mathcal{D}(P)\setminus\mathcal{D}_c$. That is  $n\beta_0 \in B_I$ for all positive integer $n$. 

Now  for every $\phi\in (\mathcal{D}(P)\setminus\mathcal{D}_c)$, $U_{\beta_0}\phi=\psi \in (\mathcal{D}(P)\setminus\mathcal{D}_c)$; $\psi$ is unique because $U_{\beta_0}$ is unitary. This implies that for every $\phi\in (\mathcal{D}(P)\setminus\mathcal{D}_c)$ there exists a unique $\psi\in (\mathcal{D}(P)\setminus\mathcal{D}_c)$ such that $\phi=U_{-\beta_0}\psi$. This further implies that $-\beta\in B_I$. Using the same argument we used above, we find that $-n\beta_0 \in B_I$ for all positive integer $n$. Thus $n\beta_0\in B_I$ for all integer $n$.

(ii) Equation (\ref{paul-Dp}) constitutes the commutator $U_{\beta'}$ in $\mathcal{D}(P)\setminus\mathcal{D}_c$, where $\beta'=n\beta_0$. $\varphi_0$  belongs to $\mathcal{D}(P)\setminus\mathcal{D}_c$ by Theorem~\ref{ccpair}. Lemma-\ref{cmlem} requires that $U_{\beta'}$ ($U_{\beta'}^{\ast}$) map $\varphi_0$ into its orthogonal complement. Thus
\begin{eqnarray*}
 \left<\varphi_m \big{|}\varphi_n \right>&=&\left<U_{\beta_0}^m \varphi_0 \big{|} U_{\beta_0}^n \varphi_0 \right>\\
&=&\left<\varphi_0 \big{|} U_{\beta_0}^{n-m} \varphi_0 \right>\\
&=&\delta_{mn}. 
 \end{eqnarray*}

\noindent That is the $\varphi_n$'s are orthonormal.  Since $U_{\beta_0}^n = U_{n \beta_0}$, theorem-\ref{comain} implies that $(P U_{n \beta_0}-U_{n \beta_0} P) \varphi_0 = n \beta_0 U_{n\beta_0} \varphi_0 \rightarrow P U_{n\beta_0} \varphi_0 = (p_0 + n\beta_0) U_{\beta_0} \varphi_0$. This and the orthonormality of the $\varphi_n$'s mean that for all integer $n$, $U_{\beta_0}^n \varphi_0$ is an eigenvector of $P$ with the eigenvalue $(p_0+n\beta_0)$.  

(iii) If $p_0$ has a multiplicity of $\lambda$, there are $\lambda$ linearly independent eigevectors corresponding to the eigenvalue $p_0$. The rest of the eigenvalues will have the same multiplicity because $U_\beta$ is a bijection. 

(iv) By (ii) some of the eigenvalues of $P$ are given by $(p_0+n\beta_0)$,  $n=0, \pm1, \pm2,\cdots$.  This implies that the eigenvalues of $P$ extend from negative to positive infinity.  The countability of the eigenvalues follows from the separability of the Hilbert space. 

(v) Let $\beta \in B_I$ but there are no $n$ and $m$ such that $\beta=p_n-p_m$.  By (ii) $U_\beta\varphi_0$ is an eigenvector of $P$ with the eigenvalue $(p_0+\beta)$, which is a contradiction.  Thus $B_I$ consists only of the eigenvalue differences of $P$. \end{proof}

\par Theorem~\ref{main} has a restatement which is important enough to be quoted as a theorem. In fact the following restatement summarizes the whole point of the paper.

\begin{theorem}[Main Theorem] \label{mainmain} Let $Q$ and $P$ be a canonical pair in some dense domain $\mathcal{D}_c$ such that equation (\ref{paul-Dc}) is satisfied. If the invariance parameter set and the point spectrum of $P$ are both not empty, then $P$ has infinitely countable eigenvalues with equal multiplicities extending from negative to positive infinity, and the invariance parameter set consists of the differences of these eigenvalues. Conversely if $P$ has semibounded or finitely countable eigenvalues then the invariance parameter set is empty. 
\end{theorem}

\begin{theorem} \label{converse} Let $Q$ and $P$ be a canonical pair where $P$ is unbounded.  Moreover, let $\mathcal{D}_a \subseteq\mathcal{D}_c$ consists of the analytic vectors of $P$ contained in $\mathcal{D}_c$ which is invariant under $Q$ and $P$.  If the restriction of $P$ in $\mathcal{D}_a$ is closed and $Q$ is bounded, then $\mathcal{D}_a$ is the trivial subspace. \end{theorem}

\begin{proof}  In ${D}_a$ the sequence of operators $V^{(n)}=\sum_{k=0}^n \frac{(-i \alpha)^k}{k!} P^k$ converges strongly to $V_\alpha$, i.e. $V^{(n)} \phi \rightarrow e^{-i \alpha P} \phi$ for all $\phi \in D_a$, for some $\abs{\alpha}<s$, the $\phi$'s being analytic vectors of the unbounded operator $P$. Also because $\mathcal{D}_a$ is invariant under $P$ and the restriction of $P$ in $\mathcal{D}_a$ is closed, $e^{-i\alpha P}\phi\in\mathcal{D}_a$. Now the invariance of $\mathcal{D}_a$ under $Q$ and $P$, and the boundedness of $Q$  ensure that $\mathcal{D}(QP^k)\supseteq\mathcal{D}_a$ and $\mathcal{D}(P^kQ)\supseteq\mathcal{D}_a$, which implies that $(QP^k-P^kQ)$ is defined in the entire $D_a$.  Since $Q$ and $P$ are a canonical pair and $\mathcal{D}_a$ is linear, it can be shown that $(QP^k-P^kQ)\phi=ikP^{k-1} \phi$ for all $\phi \in\mathcal{D}_a$.  Following the same steps employed in Theorem~\ref{comain}, we get
\begin{equation} 
	\label{pre} (Q V_\alpha^{(n)}-V_\alpha^{(n)} Q) \phi=\alpha V_\alpha^{(n-1)} \phi \end{equation}

\noindent for all $\phi \in\mathcal{D}_a$.  Because $\mathcal{D}_a$ by hypothesis is invariant under $Q$, the sequence $V_\alpha^{(n)} Q\phi$ converges strongly to $V_\alpha Q \phi$.  Also since $Q$ is bounded, thus continuous, the sequence $Q V_\alpha^{(n)} \phi$ likewise converges strongly to $Q V_\alpha \phi$.  Thus in the limit equation (\ref{pre}) reduces to
\begin{equation} \label{limit} 
	(Q V_\alpha-V_\alpha Q) \phi=\alpha V_\alpha \phi 
\end{equation}

\noindent for all $\phi \in \mathcal{D}_a$ and $\abs{\alpha}<s$.  For $\abs{\alpha}>s$, let $n$ be an integer such that $\frac{\abs{\alpha}}{n}<s$. Then for all $\phi\in\mathcal{D}_a$ $V_\alpha \phi=V_{\frac{\alpha}{n}}^n \phi \in \mathcal{D}_a$ because $V_{\frac{\alpha}{n}} \phi \in \mathcal{D}_a$. Thus for all $\phi\in \mathcal{D}_a$,
\begin{eqnarray*}
	(Q V_\alpha-V_\alpha Q)\phi&=&(Q V_{\frac{\alpha}{n}}^n-V_{\frac{\alpha}{n}}^n Q)\phi\\
		&=&nV_{\frac{\alpha}{n}}^{n-1}(Q V_{\frac{\alpha}{n}}-V_{\frac{\alpha}{n}} Q)\phi\\
		&=&\alpha V_\alpha \phi.
\end{eqnarray*}

\noindent Thus equation (\ref{limit}) holds for all $\alpha\in\Re$.

Now equation (\ref{limit}) is the commutator $C(V_\alpha,Q;D_a)$ where $C$ is the restriction of $V_\alpha$ in $\mathcal{D}_a$.  Because $V_{\alpha}\phi\in\mathcal{D}_a$ for all $\phi\in\mathcal{D}_a$ for every $\alpha\in\Re$, $V_{\alpha}$ is an invariance commutator in $\mathcal{D}_a$.  Since $V_{\alpha}$ is invertible, by Lemma~\ref{cmlem}, $V_\alpha$ and $Q$ can not be both bounded and defined everywhere.  But they are.  Therefore in order to maintain the equality in equation (\ref{limit}) $\phi$ must be the zero vector---and no other else.  Thus $\mathcal{D}_a$ is the trivial subspace. \end{proof}

\section{The Consistency Theorems Against\\Pauli's Theorem}\label{synthesis}

Theorem-\ref{ccpair} asserts that the canonical commutation relation is at most valid in a proper subspace of the Hilbert space, and in this subspace the eigenvectors of $T$ and $H$, if they exist, can not be found. Theorem-\ref{comain}.i, specifically equation (\ref{paul-Dc}) , is but the assertion of Pauli, equation (\ref{paul}); however, it is only under the conditions required by Theorem~\ref{comain}.i that equation (\ref{paul-Dc}) holds.  Pauli assumed that $\mathcal{D}_c$ contains the eigenspace of the Hamiltonian.  Under this assumption contradiction indeed arises.  But Theorem-\ref{ccpair} says otherwise---the eigenspace lies outside $\mathcal{D}_c$.  

Nevertheless, Theorem-\ref{comain}.ii allows equations (\ref{paul-Dc}) to hold within the rest of the domain of the Hamiltonian if $\mathcal{D}_c$ is dense and $\mathcal{D}(H)\setminus\mathcal{D}_c$ is invariant under $U_\beta$ for some $\beta$.  Pauli's Theorem would then hold if the invariance parameter set were the entire real line. However, Theorems-\ref{main} and \ref{mainmain} allow Theorem-\ref{comain}.ii only for Hamiltonians possessing an unbounded point spectrum extending from negative to positive infinity. Under this condition the $\beta$'s can not be arbitrary but only those eigenvalue differences of the Hamiltonian, as opposed to Pauli's assumption that $\beta$ takes an arbitrary value in $\Re$.  

Finally, Theorem-\ref{converse} implies that there is no converse of Theorem-\ref{comain} with the roles of $Q$ and $P$ interchaged in a nontrivial proper subspace of $\mathcal{H}$.  Then one can not interchange $T$ and $H$  in equation ({\ref{paul}) to contradict the assumption that the time operator is bounded. Specifically Theorem-\ref{converse} forbids one to conclude that the time operator has an unbounded spectrum spanning the entire real line.

\section{Explicit realization of the Consistency Theorems}\label{explicit}

\indent In this section we demonstrate the consistency theorems and give an explicit counter example to equation (\ref{conc}). Consider the time operator and the Hamiltonian defined below 
\begin{equation} \label{realize}
	T\psi(t)=t\psi(t),~~~~  H\psi(t)=-i\psi'(t) 
\end{equation}

\noindent in the Hilbert space $\mathcal{H}=\mathcal{L}^2[0,2\pi]$.  The domain of $T$ is the entire $\mathcal{H}$, and that of $H$ is the dense subspace
\begin{displaymath} 
	\mathcal{D}(H)=\left \{\psi(t) \in \mathcal{H}: \psi(t)\in C^{1}[0,2\pi], \psi'(t) \in \mathcal{H}, 		\psi(0)=e^{i2\pi \gamma} \psi(1), \gamma\in[0,1) \right\}, 
\end{displaymath} 

\noindent where $C^{1}[0,\pi]$ consists of all continously differentiable functions. Since the time operator is symmetric and bounded, it is self-adjoint. Also the Hamiltonian is self-adjoint as defined; moreover, it is unbounded and its eigenvectors and eigenvalues are given by $\varphi_n=(2\pi)^{\frac{-1}{2}}\exp\left[i(n+\gamma)t\right]$ and $E_n=(n+\gamma)$, $n=0, \pm1, \pm2, \cdots$, respectively. 

It is straightforward to show that the time operator and the Hamiltonian are canonically conjugate in the dense domain indicated below, i.e.
\begin{equation}\label{canoni} (TH-HT)\psi(t) = i \psi(t) \end{equation}
for all
\begin{displaymath} 
	\psi(t) \in \mathcal{D}_c=\left\{\psi(t) \in D(H): 	\psi(0)=\psi(2\pi)=0 \right \} 
\end{displaymath}
In consonance with theorem-\ref{ccpair}, the eigenvectors of the Hamiltonian lie outside $\mathcal{D}_c$ because they do not vanish at the boundaries.  It is evident that $\mathcal{D}_c$ is invariant under the action of the time operator, but not under the Hamiltonian. Since the restriction of the Hamiltonian in $\mathcal{D}_c$ is not self-adjoint and $\mathcal{D}_c$ is dense, the restriction $H|_{D_c}$ is closed. Because all conditions of Theorem~\ref{comain}.i are satisfied by the time operator, the Hamiltonian, and the domain $\mathcal{D}_c$, then for all $\psi \in \mathcal{D}_c$ and $\beta \in \Re$, we have 
\begin{equation} \label{exm} 
	(HU_\beta-U_\beta H) \psi(t)=\beta U_\beta \psi(t) 
\end{equation}
where $U_\beta=\exp(-i\beta T)$, which can easily be verified.  While equation (\ref{exm}) is valid for every real number $\beta$, it does not imply equation (\ref{conc}) because the eigenvectors of  the Hamiltonian lie outside $\mathcal{D}_c$. 

Equation (\ref{exm}) holds in $\mathcal{D}(H)\setminus\mathcal{D}_c$ if there are values of $\beta$ such that $\mathcal{D}(H)\setminus\mathcal{D}_c$ is invariant under the action of $U_\beta$, or the invariance parameter set is not empty.  Now $B_I$ consists of those points $\beta \in \Re$ such that for all $\psi \in \mathcal{D}(H) \setminus\mathcal{D}_c$, $U_\beta \psi \in \mathcal{D}(H)$.  Let $\psi \in \mathcal{D}(H)$.  The boundary condition requires $(U_\beta \psi)(0)=e^{i2\pi \gamma} (U_\beta \psi)(2\pi)$, which gives $\psi(0)=e^{-i2\pi \beta} e^{i2 \pi \gamma} \psi(2 \pi)$. Because $\psi(0)=e^{i2\pi\gamma} \psi(2\pi)$, we have $1=\exp(i2\pi \beta)$.  Thus $B_I$ consists of the points $\{\beta_n=n, n=0, \pm1, \pm2, \cdots \}$ and only these points, which, by inspection, are just the eigenvalue differences of the Hamiltonian. Since the invariance parameter set is not empty, then for all $\psi \in \mathcal{D}(H)\setminus\mathcal{D}_c$ and $\beta \in B_I$, equation (\ref{exm}) holds.  For non-integer values of $\beta$, $U_\beta$ maps all $\psi \in \mathcal{D}(H) \setminus\mathcal{D}_c$ outside of the domain of $H$ so that $HU_{\beta}\varphi$ is not defined in the entire domain of the Hamiltonian, and for this case commutation relation (\ref{exm}) does not hold. One can not then appeal to equation (\ref{exm}) to arrive at the conclusion that equation (\ref{canoni}) implies that the spectrum $H$ is the real line.

Now there is a converse of equation (\ref{exm}) (for all $\alpha\in\Re$) , i.e. the roles of $H$ and $T$ interchanged, if  there exists a subspace $\mathcal{D}_a\subseteq \mathcal{D}_c$ consisting of analytic vectors of $H$ and the restriction of $H$ in $\mathcal{D}_a$ is closed. The vectors in $\mathcal{D}_c$ which are infinitely differentiable complex valued functions with supports away from $0$ and $2\pi$ are analytic vectors of $H$. Thus $\mathcal{D}_a$ exists. However, the restriction of $H$ in $\mathcal{D}_a$ is not closed because while $\psi_n=\sum_{k=0}^n \frac{(-i\alpha)^k}{k!} H^k\phi \in \mathcal{D}_a$ for all $n$ and $\alpha\in\Re$ and for every $\phi \in \mathcal{D}_a$ the limit of $\psi_n$ as $n$ approaches infinity does not necessarily exists, i.e. the vector limit does not belong to the Hilbert space. Nevertheless, for sufficiently small $\alpha$ it is straightforward to show that
\begin{equation}\label{cow}
(TV_{\alpha}-V_{\alpha}T)\phi(q)=\alpha V_{\alpha}\phi(q)
\end{equation}
for some $\phi\in\mathcal{D}_a$ where $V_{\alpha}=e^{-i\alpha H}$. If we insist that equation (\ref{cow}) must hold for all $\alpha\in\Re$, then we have to restrict $\mathcal{D}_a$ to the zero vector alone because this is the only vector that can be translated without bring it out of the interval $[0,2\pi]$. This means that we can not interchange the roles of $H$ and $T$ to infer that the spectrum of $T$ is the entire real line.

\section{The particle in a box}\label{free}

So it is consistent to assume the existence of a bounded, self-adjoint time operator. But under what circumstances that a bounded, self-adjoint and canonical time operator can arise in standard quantum mechanics? In this section we demonstrate how quantization of a classical time observable can give rise to a bounded and self-adjoint time operator canonically conjugate to the corresponding Hamiltonian. However, we will not delve into elaborating the physical content of the constructed time operator beyond the result that it is a consequence of quantization, doing so is beyond the scope of this paper. Instead we will deal with it elsewhere (Galapon \& Bahague 2001 in preparation).

Let us consider a particle of unit mass confined within a box of two units of length. We assume that the particle is subject to the potential $V(q)=0$ for $-1\leq q\leq +1$ and $V(q)=\infty$ elsewhere. Classically the Hamiltonian between the boundaries is given by $H_c=\frac{1}{2}p^2$. Let $q$ be the position of the particle. If $\left|q\right|\leq 1$ and if $p\neq 0$ is the momentum of the particle, then the time that the particle will arrive (first arrival time, i.e. without reflection from the boundaries) at the origin is $T_c=- qp^{-1}$. We note that the pair $(H_c,T_c)$ is a canonical pair, i.e. $\{H_c,T_c\}=1$. We ask, Is there a quantization of the pair $(H_c,T_c)$ such that both operators are self-adjoint and are quantum canonical pair in some non-trivial subspace of the Hilbert space? The answer is, Yes. 

Let us attach the Hilbert space $\mathcal{H}=L^2[-1,1]$ to the system, the space of all Lebesgue square integrable functions over the closed interval $[-1,1]$. We assume the following self-adjoint quantizations of the position and the momentum operators: The position operator $\opr{q}$ is the multiplication by $q$ operator whose domain $\dom{\opr{q}}$ is the entire Hilbert space. The momentum operator is the operator $\opr{p}=-i \partial_{q}$ whose domain is $\dom{p}=\left\{\phi\in\mcal{H}: \phi(q)\, \mbox{absolutely continuous} ,\,\phi'(q)\in\mcal{H},\,\phi(-1)=\phi(1)\right\}$. The domain is so choosen such that the Hamiltonian is purely kinetic, i.e. $\opr{H}=\frac{1}{2}\opr{p}^2$. Explicitly the Hamiltonian is given by
\begin{equation}\label{freeham}
\left(\opr{H}\varphi\right)\!(q)=-\frac{1}{2}\frac{d^2\varphi}{dq^2}\;\mbox{for all}\;\varphi(q)\in\mathcal{D}(H)
\end{equation}
where its domain is
\begin{equation}
\mathcal{D}(\opr{H)}=\left\{\varphi\in \dom{\opr{p}}: \varphi'\, a.c.,\, \varphi''\in \mathcal{H},\,\varphi'(-1)=\varphi'(1)\right\}
\end{equation}
Because the momentum operator is self-adjoint, the Hamiltonian is likewise self-adjoint. The common eigenvectors of $\opr{p}$ and $\opr{H}$ are $\phi_n=2^{-\frac{1}{2}}\exp(i\,n\pi\,q)$, and their respective eigenvalues are $p_n=n\pi$ and $E_n=n^2\pi^2/2$, for all $n=0,\pm1,\pm2,\cdots$. Evidently the Hamiltonian is discrete and semibounded. According to Pauli's argument, no self-adjoint time operator can be constructed for such a Hamiltonian. However, we will show below that a quantization of the time of arrival is self-adjoint and at the same time canonically conjugate to the quantized Hamiltonian.

Now let us quantize the time of arrival $T_c=-qp^{-1}$. Since $\opr{q}$ and $\opr{p}$ do not commute, we have to choose a particular ordering scheme. We choose symmetric ordering, and we get the formal time of arrival operator $\opr{T}=-\frac{1}{2}\left(\opr{q}\opr{p^{-1}}+\opr{p^{-1}}\opr{q}\right)$. Of course, $\opr{T}$ make sense only if the operators involved are well defined. Since $\opr{q}$ appears in first power in $\opr{T}$, $\opr{T}$ is well defined if $\opr{p^{-1}}$ is defined in $\mathcal{H}$. However $\opr{p^{-1}}$ is not defined because $\opr{p}$ has no inverse, the zero being an eigenvalue of $\opr{p}$. But this can be remedied. The pathology arises from the one-dimensional subspace spanned by the state of vanishing momentum, the null subspace $\mcal{N}(\opr{p})$. But this subspace has no bearing to the problem because the question when a given particle arrives makes sense only when the particle is in motion. We expect then that $\opr{T}$ is well defined if the contribution of the null subspace is removed. Technically this can be accomplished as follows: Let $\mcal{P}$ and $\mcal{P}^{\perp}$ be the projections unto the closures of the subspaces $\mcal{N}(\opr{p})$ (the subspace spanned by the zero momentum state) and $\mcal{N}(\opr{p})^{\perp}$ (the subspace spanned by the non-vanishing momentum states), respectively. Since $\mcal{N}(\opr{p})$ is invariant under $\opr{p}$ (that is, $\opr{p}\mcal{N}(\opr{p})=\{0\}\subset \mcal{N}(\opr{p})$), $\mcal{N}(\opr{p})$ and $\mcal{N}(\opr{p})^{\perp}$ reduce $\opr{p}$. Because $\opr{p}$ is self-adjoint, its rectrictions on $\mcal{N}(p)$ and $\mcal{N}(p)^{\perp}$ ($\opr{p}_{\mcal{N}}: \mcal{P}\mcal{D}(\opr{P})\subseteq \mcal{P}\mcal{H}$ and $\opr{p}_{\mcal{N}^{\perp}}: \mcal{P}^{\perp}\mcal{D}(\opr{p})\subseteq \mcal{P}^{\perp}\mcal{H}$)  are both self-adjoint. The restriction $\opr{p}_{\mcal{N}^{\perp}}$ has trivial null-space, so that its inverse, $\opr{p}_{\mcal{N}^{\perp}}^{-1}$, exists in $\mcal{P}^{\perp}\mcal{H}$. But this inverse exists only in $\mcal{P}^{\perp}\mcal{H}$ and not in $\mcal{H}$. This can be addressed by extending 
$\opr{p}_{\mcal{N}^{\perp}}^{-1}$ in the entire $\mcal{H}$. First we note that $\opr{p}_ {\mcal{N}^{\perp}}^{-1}$ is self-adjoint and bounded. It can be shown that  $\opr{p}_{\mcal{N}^{\perp}}^{-1}$ is subnormal and it admits a unique minimal  extension in the entire $\mcal{H}$. Its minimal extension is the bounded and self-adjoint operator $\opr{P}^{-1}=\mcal{P} \opr{p}_{\mcal{N}^{\perp}}^{-1}\mcal{P}$. This operator can be interpreted as the quantization of the classical obervable $p^{-1}$ for $p\neq 0$. Substituting $\opr{P^{-1}}$ for $\opr{p^{-1}}$ in $\opr{T}$, we get the operator $\opr{T}=-\frac{1}{2}\left(\opr{q}\opr{P}^{-1}+\opr{P}^{-1}\opr{q}\right)$, which is {\it a} quantization of $T_c$. Note that both $\opr{q}$ and $\opr{P^{-1}}$ are bounded, everywhere defined, self-adjoint operators. Since $\opr{T}$ is symmetric under the exchange of $\opr{q}$ and $\opr{P^{-1}}$, it is likewise bounded, everywhere defined and self-adjoint.

So $\opr{T}$ is self-adjoint, but is it canonically conjugate to the Hamiltonian according to equation (\ref{ccr})? Yes. To see this we need to find the coordinate representation of $\opr{T}$. Since $\opr{q}$ is already diagonal in the Hilbert space, we need only to write the coordinate representation of $\opr{P^{-1}}$. Its representation is 	$(\opr{P^{-1}}\varphi)\!(q)=\pi^{-1} {\,\,\,}^{\prime}{\sum_{n=-\infty}^{\infty}}(\phi_n,\varphi)n^{-1} \phi_n(q)$ for all $\varphi(q)\in\mcal{H}$, where the prime indicates summation without the contribution of $n=0$ and the $\varphi_n$'s are the eigenfunctions of the momentum operator. This leads to the integral operator representation of $\opr{T}$,
\begin{equation} \label{fredholm}
	\left(\opr{T}\varphi\right)\!(q)=\int_{-1}^{+1}T(q,q')\,\varphi(q')\,dq' 		\;\;\;\mbox{for all}\;\;\;\varphi(q)\in\mcal{H},
\end{equation}
where the kernel is given by
\begin{equation} \label{periodic}
	T(q,q')=\frac{1}{4i}(q+q')\mbox{sgn}(q-q')-\frac{1}{4i}(q^2 -{q'}^2).
\end{equation}
Using equations (\ref{fredholm}) and (\ref{periodic}), it is straigthforward to show that the Hamiltonian and the quantized time of arrival operator are canonically conjugate in a non-trivial subspace of the Hilbert space, i.e.
\begin{equation}
\left(\left(TH-HT\right)\varphi\right)\!(q)=i\,\varphi(q)\;\mbox{for all}\;\varphi(q)\in\mathcal{D}_c
\end{equation}
where
\begin{equation}
\mathcal{D}_c=\left\{\varphi\in\mathcal{H}:\int_{-1}^{1}q^k\varphi(q)\,dq=0,\,k=0,1; \varphi(\partial)=0,\varphi'(\partial)=0\right\}\nonumber
\end{equation}
We note that $\mathcal{D}_c$ is orthogonal to the two dimensional subspace $\left\{\phi=a+bq,\, a,b\in \mcal{C}\right\}$; thus $\mathcal{D}_c$ is not dense but closed. 

We asked whether the classical canonical pair $(H_c,T_c)$ has a self-adjoint and canonical quantization. We have demonstrated above that it has, contrary to Pauli's claim. How could the pair $(\opr{H},\opr{T})$ be self-adjoint and canonical without contradiction? First, eventhough $\opr{H}$ and $\opr{T}$ satisfy (\ref{ccrel}), one can not proceed to equation (\ref{paul}) in the entire $\mathcal{D}_c$, because the canonical domain $\mathcal{D}_c$ is not invariant under the action of the time operator; for example, $ \varphi(q)=\cos(\pi q)+\cos(2\pi q) \in \mathcal{D}_c$ but $\left(T\varphi\right)\!(q)\neq \mathcal{D}_c$. This means that eventhough $T$ is bounded such that the expansion $\exp\left(-i\beta T\right)=\sum_{k=0}^{\infty}\frac{(-i\beta)^k}{k!}T^k$ holds everywhere, we will not arrive at the right hand side of equation (\ref{paul}) because $\left(T^k H- HT^k\right)\varphi\neq i\hbar k T^{k-1}\varphi$ for all $\varphi\in\mathcal{D}_c$. To see this consider $\left(T^2 H-HT^2\right)$ in $\mathcal{D}_c$. Using the linearity of  $\mathcal{D}_c$ and the fact that $\mathcal{D}_c\subseteq\mathcal{D}(TH)\cap\mathcal{D}(HT)$, we have $\left(T^2 H-HT^2\right)\varphi=\left(T(TH-HT)+(TH-HT)T\right)\varphi$. While $\left(T(TH-HT)\right)\varphi=i\hbar\, T\varphi$ for all $\varphi\in\mathcal{D}_c$, $\left((TH-HT)T\right)\varphi\neq i\hbar\, T \varphi$ because $\mathcal{D}_c$ is not invariant under $T$, i.e. $T\varphi\not\in \mathcal{D}_c$.  We draw the attention of the reader that this also means that the pair do not satisfy equation (\ref{paul-Dc}); and, consequently, they do not satisfy equation (\ref{paul-Dp}) also. This suggests that the canonical pair considered in Section (\ref{explicit}) and the canonical pair in this Section belong to two different classes of solution to the canonical commutation relation (\ref{ccr}). We will have more to say about this observation below.

\section{Why Pauli's theorem has stood up to this day}\label{discussion}

In this paper we have explicitly demonstrated that the sweeping generalization of Pauli's Theorem is not justified. It is natural to ask why Pauli's argument has stood to this day despite some well-known examples of canonical pairs that do not possess spectra spanning the entire real line. 

One cause of the oversight is that all of the examples---the momentum and the position operators of a particle trapped in a box, the angular momentum and the angle operators, the harmonic oscillator number and phase operators---appear in contexts different from the context of  any problem involving dynamical concept of time (except in Garrison and Wong (1970) where a quantum clock is constructed out of the self-adjoint phase operator). This is confounded by the utter silence of the authors of the cited works in relation to Pauli's well known claim. A related cause is the deeply rooted conviction that physical time, if ever it has to be represented by an operator and whatever the modifier \textsl{physical} may mean, must be an unbounded operator. In the above examples, it is reasonable for the position operator to be bounded because the particle is confined in the first place. Likewise, it is reasonable for the angular and the phase operators to be bounded because the azimuthal angle and the phase are restricted within the $2\pi$-window. But for a dynamical time problem such as the time of arrival problem, the time operator, if ever it exists, has to be unbounded. This conviction, on a closer look, is another spin-off  from Pauli's Theorem: The converse of equation (\ref{conc}) derived by interchanging the roles of the Hamiltonian and the time operator implies that the time operator must have unbounded and continuous spectrum in both directions, i.e. $T$ is unbounded. However, if we take a pragmatic stand on the issue, no feasible experiment (at least for now) can decide if $T$ is indeed unbounded because experiments have to be performed within a finite interval of time. Any experiment designed to span the entire spectrum of an unbounded $T$ would have to run indefinitely and outlive the experimenter and his succesors. For this reason it is not difficult to see that a bounded $T$ is physically more realizable than an unbounded one. The boundedness of $T$ would indicate the occurence of every expected event. Any unbounded operator representation of time is then an idealization incorporating the none occurence of the expected event, e.g. a particle does not arrive or a nucleon does not decay at all.

Another seemingly corroborative result demonstrating Pauli's claim is the overwhelming pessimistic conclusion of Allcock (1969) that the free motion time of arrival problem can not be accomodated within the standard single Hilbert space formulation of quantum mechanics. Allcock's result has been construed by some as a generalization of Pauli's Theorem.  Allcock's conclusion rests on the non-self-adjointness and the lack of any self-adjoint extension of the free motion time of arrival operator, $T=(qp^{-1}+p^{-1}q)2^{-1}$, in free unbounded space, and on heuristic arguments supported by two postulates. While the former is a valid argument based on the standard axiom of quantum mechanics (that observables are represented by self-adjoint operators), the latter is at best a conjecture whose conclusion can not be taken at face value without firmly establishing the generality of the assumed conditions in the postulates. What can we say about the non-self-adjointness of $T$? Does it constitute a proof to Pauli's claim? Of course not. In fact the non-self-adjointness of $T$ has nothing to do Pauli's theorem.  More importantly, we point out that the non-self-adjointness of $T$ can be dealth with by confinement as we have demonstrated above.

Another reason of the oversight is the existence of examples that appear to corroborate the conclusion of Pauli. If we apply the reasoning leading to equation (\ref{conc}) to the position and the momentum operators in free unbounded space (which are both unbounded and canonically conjugate in a dense subspace of $L^2 (\Re)$ in which they are essentially self-adjoint) we are led to conclude that the spectra of both operators span the entire real line. This conclusion is undeniably correct. On the other hand, if we apply the same reasoning to the position and the momentum operators in the half line (which are both unbounded and canonically conjugate in a dense subspace of $L^2(\Re^{+})$ in which only the position operator is essentailly self-adjoint) we find that according to Pauli's argument the momentum operator can not be self-adjoint because the position operator is bounded from below. This conclusion is again undeniably correct. These examples have in fact actually become textbook examples in the discussion of the spectral properties of canonical pairs (Cohen-Tannoudji 1977; Gottfried 1966). But  these examples are mere coincidences. The argument leading to equation (\ref{conc}) is too formal that it can not be blindly applied to unbounded operators like the momentum operator without abandoning the deceptively simple assumptions of Pauli. Specifically, in the two foregoing examples, one has to contend with Pauli's assumption on the existence of the eigenvectors of the canonical pair against the fact that the position and the momentum operators in full and half lines do not have eigenvectors. Of course, one can appeal to the existence of generalized eigenvectors of the operators in full or half line; but this can not be done without further ramifications to justify the step leading to equation (\ref{conc}).  

Perhaps most convincing of all is the existence of a ``rigorous" proof and a generalization of Pauli's Theorem (Srinivas \& Vijayalakshmi 1981). However, we point out that  what was proven and generalized is not actually Pauli's Theorem but something else. The whole proof of Srinivas and Vijayalakshmi rests on the following well known result: Let $Q$ and $P$ be the respective generators of the strongly continuos one parameter unitary groups $U_{\alpha}=\mbox{exp}(i\alpha P)$ and $V_{\beta}=\mbox{exp}(i\beta Q)$, where $\alpha, \beta \in \Re$.  Let $\Delta$ be a Borel subset of $\Re$; let $\Delta\rightarrow E^Q (\Delta)$ and $\Delta\rightarrow E^P (\Delta)$ be the respective unique projection valued measures corresponding to $Q$ and $P$. If $Q$ and $P$ form a system of imprimitivities based on the real line $\Re$, i.e.
\begin{eqnarray}\label{rel1}
E^Q (\Delta+\alpha)=U_{\alpha}^{-1} E^Q (\Delta) U_{\alpha},\\
E^P (\Delta+\beta)=V_{\beta} E^P (\Delta) V_{\beta}^{-1}\label{rel11},
\end{eqnarray}
then
\begin{eqnarray}\label{rel2}
U_{\alpha}V_{\beta}&=&\mbox{e}^{i\alpha\beta} V_{\beta}U_{\alpha}\\
V_{\beta}^{-1}PV_{\beta}\phi&=&\left(P+\beta I\right)\phi\label{rel21}\\
U_{\alpha}^{-1}QU_{\alpha}\phi&=&\left(Q-\alpha I\right)\varphi \label{rel3}
\end{eqnarray}
for all $\phi\in\mcal{D}(P)$ and $\varphi\in\mcal{D}(P)$, and for all $\alpha,\beta\in\Re$; moreover, there exists a common dense subspace $\mcal{D}_c$ of $\mcal{D}(P)$ and $\mcal{D}(Q)$ such that
\begin{equation}\label{ccc}
(QP-PQ)\psi=i\psi,
\end{equation}
for all $\psi\in\mcal{D}_c$. 

Quoting the equivalence of equations (\ref{rel1})-(\ref{rel3}), Srinivas and Vijayalakshmi restated Pauli's Theorem in the following way:
\begin{quote}
If the self-adjoint operator $H$ is semibounded, then there does not exist a self-adjoint operator [$T=\int_{-\infty}^{\infty}t\,dE^T$ such that the relation
\begin{equation}\label{srinivas}
\mbox{exp}\left(iH \alpha\right)E^T(\Delta)\mbox{exp}\left(-iH \alpha\right)=E^T(\Delta+\alpha)
\end{equation}
is satisfied for all $\alpha\in\Re$].
\end{quote}
The above assertion is correct and it is a consequence of equations (\ref{rel1}) and (\ref{rel11}) (with $Q=H$ and $P=T$). However, it is {\it not} Pauli's Theorem: Pauli's Theorem, in its barest essence, asserts that the canonical commutation relation (\ref{ccc}) implies equations (\ref{rel1}) and (\ref{rel11}) through the formal manipulation presented in Section-\ref{revisit}. To assert the equivalence of Pauli's Theorem and Srinivas and Vijayalakshmi's restatement is to assert the equivalence of the canonical commutation relation and the imprimitivity relation (\ref{rel1}). Of course this assertion is not correct. Srinivas and Vijayalakshmi even acknowledged this.  Their restatement and generalization then can not be Pauli's Theorem. So where is the confusion? While they acknowledged that equations (\ref{rel1}) and (\ref{rel11}) imply equation (\ref{ccc}) but not vice versa, they apparently assumed that equations (\ref{rel21}) and (\ref{rel3}) are universal consequences of the canonical commutation relation (\ref{ccc}). With this assumption, a connection between Pauli's Theorem and their restatement is apparently established: Comparing equations (\ref{rel21}) and (\ref{rel3}) with equation (\ref{conc}), we find that they are ``equivalent." However, their equivalence is only formal, i.e. without regard to the domains of the operators involved---again. It is true that equations (\ref{rel21}) and (\ref{rel3}) hold in the entire domains of $Q$ and $P$, but this is a consequence of equation (\ref{rel1}). Without (\ref{rel1}), commutation relations (\ref{rel21}) and (\ref{rel3}) may still hold but now only in a proper subspace of the domains of $Q$ and $P$. This is the case considered in Section-\ref{explicit}.

\section{Quantum Canonical Pairs}\label{canonical}

Pauli's theorem has been asserting that a pair of self-adjoint operators, $Q$ and $P$, satisfying the canonical commutation relation (\ref{defccp}) form a system of imprimitivities based on the real line, i.e. both operators satisfying the commutation relations (\ref{rel1}) and (\ref{rel11}). This has given the impression that inorder for a canonical pair to be meaningful, the pair must satisfy the imprimitivity relation. Moreover, Pauli's theorem has led many to believe that the canonical commutation relation (\ref{ccr}) only admits solutions that form a system of imprimitivities on $\Re$, so that any operator canonically conjugate to a semibounded Hamiltonian must necessarily be non-self-adjoint (for example Srinivas \& Vijayalakshmi 1981, Giannitrapani 1997, Park 1984, Delgado \& Muga 1997, Olhovsky \&Recami (1974), Toller 1997,1999, Cohen-Tannoudji 1977, Gottfried 1966).  Our examples above and the examples of the much ignored earlier works of Galindo (1984), Garrison and Wong (1970), Segal (1967), and especially Nelson (1959) demonstrate otherwise, that in fact there are numerous solutions to equation (\ref{ccr}). Pauli's Theorem has been requiring solutions for $T$ and $H$, regardless of the interpretation for $T$, that form a system of imprimitivities based on the real line under the guidance of the erronous logic leading to equation (\ref{conc}). Consequently Pauli's Theorem has brushed aside and downplayed the rest of the solution set of the canonical commutation relation for a given Hamiltonian, thus, ignoring their possible significant physical contents. 

At this point, one may raise the question, If we accomodate solutions to equation (\ref{ccr}) beyond Pauli's solution in standard quantum mechanics, what physical relevance can we attach to a canonical pair without the imprimitivity requirement?  Let us answer the question by refering to the well-known position and momentum operators, $(\opr{q},\opr{p})$, in the entire real line $\Re$ and in a bounded segment of $\Re$.  In free space $\Re$, it is well known that the pair $(\opr{q},\opr{p})$ are self-adjoint, and form a canonical pair and a system of imprimitivities based on $\Re$. The pair satisfies these properties because of (i) the fundamental axiom of quantum mechanics that the propositions for the location of an elementary particle in different volume elements are compatible, and from (ii) the fundamental homogeneity of free space, i.e. points in $\Re$ are indistinguishable (Mackey 1968, Jauch 1968). The former (i) naturally leads to the self-adjoint position operator $\opr{q}$ in $\Re$; while the later (ii) requires the existence of a unitary operator generated by the momentum operator such that the projection valued measure (PV) of $\opr{q}$ satisfy equation (\ref{rel1}) (with $Q=\opr{q}$); in fact equation (\ref{rel1}) is the exact mathematical statement of homogeniety of free space. Then symmetry dictates that the PV measure of $\opr{p}$ must satisfy equation (\ref{rel11}) (with $P=\opr{p}$). Now equation (\ref{rel1}), together with the fact that $U_{\alpha}$  and $V_{\beta}$ form a representation of the additive group of real numbers, leads to the well known Weyl's commutation relation (\ref{rel2}). This relation finally implies the canonical commutation relation $(\opr{qp}-\opr{pq})\subset i\hbar\,\opr{I}$ enjoyed by $\opr{q}$ and $\opr{p}$. In a bounded segment of $\Re$, say the segment $[-1,1]$, the pair $(\opr{q},\opr{p})$ are still self-adjoint (for example, the position and momentum pair in Section-\ref{free}) and form a canonical pair. However, they now fail to form a system of imprimitivities based on $\Re$.  This non-imprimitivity (on $\Re$) is a consequence of the fact that the segment $[-1,1]$ is not homogenous; that is, points in the spatial space available to the particle are distinguishable, the boundaries being the distingushing factor. For example one point can be nearer to the left boundary than another point. Thus equation (\ref{rel1}) can not be imposed upon the position operator in $[-1,1]$, because it is the statement of the homogeneity of free space in $\Re$. Imposing equation (\ref{rel1}) on the position operator in a bounded segment of $\Re$ is imposing homogeneity in an intrinsicaly inhomegenous space. 

The above two pairs of canonical operators represent two distinct solutions of equation (\ref{ccr}). They are distinct because they have different spectral properties. Should we discriminate one from the other? If we impose imprimitivity based on $\Re$ on all acceptable canonical pairs, then the position and the momentum operators in a bounded segment of $\Re$ are unacceptable canonical pairs, because they don't satisfy the imprimitivity requirement. However, we all do accept this pair. And it is not difficult to understand why. As what we have discussed above, the imprimitivity of the position and the momentum operators in $\Re$ is a consequence of the symmetry possessed by the configuration space of the particle---the homogeniety of free space or Euclidean invarince of $\Re$. Also the non-imprimitivity on $\Re$ of the ``same" operators in a bounded segment of $\Re$ is a consequence of the boundedness and inhomogeneity of the configuration space available to the particle.  Thus imprimitivity based on the real line of a canonical pair is not an esteemed property that has to be imposed on all canonical pairs. That is we can not prefer one solution of (\ref{ccr}) from other solutions without considering the physical context of the solution being sought, a point earlier indicated by Garrison and Wong (1970). 

With the example of the position and the momentum operators above, it is clear that we can not separate the distinct solutions of the canonical commutation relation from some underlying principles. That is the set of properties of a specific solution to equation (\ref{ccr}) is consequent to a set of underlying fundamental properties of the system under consideration or to the basic definitions of the operators involved or to some fundamental axioms of the theory or to some postulated properties of the physical universe; this is to say that a specific solution to the canonical commutation relation is canonical in some sense. It is concievable to impose that a given pair be canonical as a priori requirement based, say, from its classical counterpart, but not the sense the pair is canonical without a deeper insight, say,  into the underlying properties of the system. In other words, we don't impose in what sense a pair is canonical if we don't know much, we derive in what sense instead. Furthermore, if a given pair is known to be canonical in some sense, then we can learn more about the system or the pair by studying the structure of the sense the pair is canonical. Thus it is natural to expand the class of physically acceptable canonical pairs (in standard quantum mechanics) to include any given pair of densely defined, self-adjoint operators, (\opr{Q},\opr{P}), in a separable, infinite dimensional Hilbert space, $\mcal{H}$, satisfying the canonical commutation relation in some nontrivial, proper dense or closed  subspace of $\mcal{H}$, with the assumption that the properties of these pairs can be meaningfully anchored to some underlying principles.

With the foregoing assertion, we can not be adamant with the idea of introducing self-adjoint time operators not satisfying the imprimitivity requirement.  It is widely acknowledged that time as an observable is multi-faceted (see for example Busch 1990$a$).  It makes more sense then to anticipate that these different facets do not necessarily demand similar sets of requirements on time operators as with the position and the momentum operators. Thus it may not be necessary in some cases to require imprimitivity. For example, self-adjoint time operators may enter in the construction of ideal quantum clocks in which imprimitivity is not explicitly mandatory. An ideal quantum clock requires a one-parameter family of states, $\Lambda_{\tau}$, with the property $e^{-i t H}\Lambda_{\tau}=\Lambda_{\tau+t}$, and a self-adjoint operator $T$, whose measurement on $\Lambda_{\tau}$ yields $\tau$ with negligible dispersion (Susskind \&Glogower 1968). An explicit example of such is the oscillator clock of Garrison and Wong (1970) in which a bounded and self-adjoint time operator (canonically conjugate to the oscillator Hamiltonian) is used for $T$. Another example may arise from the quantization of classical observables, which are real valued functions of the classical canonical coordinates. The classical time of arrival, for example, is a legitimate classical observable by virtue of its dependence on the canonical coordinates. And quantization of this may lead to a legitimate quantum observable. An explicit example of this is the self-adjoint time of arrival operator we have constructed in Section-\ref{free}.

But what if we require imprimitivity on $\Re$? With the discussion above on the different solutions to the canonical commutation relation, we are free to impose imprimitivity as long as we understand the physical underpinnings of the required property; equivalently, we can not blindly impose imprimitivity without understanding why we are requiring the property. But for semi-bounded Hamiltonians no self-adjoint time operator would exist to satisfy the requirement (but this does not imply that a self-adjoint time operator canonically conjugate to the Hamiltonian does not exist as what we have been discussing all along).  In the course of the history of the quantum time problem, it has been suggested that covariance should replace the imprimitivity requirement (under the false impression that the canonical commutation and the imprimitivity relations are equivalent). This would lead to covariant non-self-adjoint time operators canonically conjugate to semibounded Hamiltonians, and these operators are treated as positive-operator-valued-measure (POVM) observables in the non-standard quantum mechanics. If one upholds the legitimacy of extending quantum obervables to POVM to accommodate non-self-adjoint observables, then covariant non-self-adjoint time operators are just one class and not the only class of solutions to the canonical commutation relation for a given Hamiltonian. Covariance can not be exclusively imposed upon time operators in order to be meaningful. Covariance can be seen as a specific property of one class of solution to the canonical commutation relation that can be anchored on specific problems, such as continuous measurements of occurrence of events (Srinivas \& Vijayalakshmi 1981). We note though that such solutions may be justified on physical grounds as long as they can be shown consistent with the axioms of quantum mechanics. We mention that over requiring covariance can lead to non-normalizable positive operator valued measures (Srinivas \& Vijayalakshmi 1981). Now this is an example when a solution to the canonical commutation relation is in conflict with the axioms of quantum mechanics; and thought is required to consider whether they are acceptable or not, an acceptance of which requires further revision of the axioms of quantum mechanics.

\section{Conclusion}\label{conclusion}

Since Pauli concluded that ``the introduction of time operator [in standard quantum mechanics] must be fundamentally abandoned", it has been tacitly understood by the majority that any effort to incorporate a dynamical quantum theory of time has to be undertaken outside the standard formulation of quantum mechanics. And indeed there has been no shortage of non-standard solution to the different aspects of the quantum time problem, solutions ranging from introducing dynamical time operators as non-self-adjoint maximally symmetric operators (Egusquiza \& Muga 1999; Giannitrapani 1997; Busch {\it et. al} 1994; Holevo 1978; Helstrom 1976, 1970) to non-self-adjoint non-symmetric operators (Haba \& Novicki 1976) to solving the problem in another quantum mechanical platforms (Eisenberg \& Horwitz 1997; Halliwell \& Zafiris 1997; Blanchard \& Jadczyk 1996; Holland 1993). While the aforementioned solutions may have merits on their own and may have shed light on some aspects of the problem (for example the extension of quantum mechanics to include POV's as legitimate quantum observables), it is important to realize that much of their introduction have been motivated by the no-go theorem of Pauli, and, thus, we might have missed some important aspects of the problem in the process of introducing these solutions. 

What could have we missed? It has been generally believed that no self-adjoint time operator exists in the standard formulation of quantum mechanics, again, because of Pauli's theorem. But in this paper, we have explicitly shown that this belief is unfounded, that in fact there may exists a class of self-adjoint time operators canonically conjugate to a given semibounded Hamiltonian. In the course of the development of quantum mechanics, these self-adjoint time operators have been thought non-existent---and thus have been neglected. If Pauli's theorem has motivated us in the past to seek solutions beyond the standard quantum mechanics, the work herein reported should motivate us to go back to the standard formulation, not to obviate what earlier efforts have already achieved, but to give due attention to what has long been neglected. A specific direction to head off is to pursue the investigation of the whole class of solutions to the canonical commutation relation in the standard formulation, the relationship of the different solutions to first physical principles (e.g. homogeneity of free space leads to the solution of the pair of position and momentum operators in $\Re$ (Jauch 1968; Mackey 1968)), and their relationship to specific problems (Garrison \& Wong 1970).  Now could it be possible that the reason that there is still so much controversy in the quantum time problem because the entirety of its picture is not yet fully understood? It is concievable that understanding the physical underpinnings of the different solutions of the canonical commutation relation can give us a better picture of the quantum time problem and may bring us nearer to a consensus.

\section*{Acknowledgements}  The author acknowledges the generosity of L.P. Horwitz in answering some questions during the initial stages of this work.  Likewise, to G. Muga for communicating references (Toller 1997, 1999), and his encouraging interest in this work; to P. Busch for sending the author copies of his works which are otherwise inaccesible to the author; to I. L. Egusquiza for his careful reading of and pointing out vague statements in the original manuscript; to R. Ramos and S. Atienza for their assistance in collecting many of the references herein cited. The author is grateful to M. Navarro and N. Reyes of UP-Mathematics Department in their carefull reading of the initial draft of this paper, which resulted in the restatement of some of the theorems and proofs; however, the author is still responsible for whatever remaining shortcomings this paper may have.

\label{lastpage}
\end{document}